# The Genetic Code Boolean Lattice

Robersy Sánchez [1 3*], Eberto Morgado [2], Ricardo Grau [2 3]

[1] Research Institute of Tropical Roots, Tuber Crops and Banana (INIVIT). Biotechnology group. Santo Domingo. Villa Clara. Cuba.

[2] Faculty of Mathematics Physics and Computation, Central University of Las Villas, Villa Clara, Cuba.

[3] Center of Studies on Informatics, Central University of Las Villas, Villa Clara, Cuba

## ABSTRACT

The algebraic structures of the genetic code are most important to obtain additional information about the semantic code and its applications. In this paper we define two dual Boolean codon lattices of the genetic code using hydrogen bond numbers and the chemical types of bases: purines and pyrimidines. The Boolean lattices reflect the role of hydrophobicity in the distribution of codon assignments to each amino acid. Particularly, the symmetric images of codons with adenine as second base coding to hydrophilic amino acids are always codons with uracil as second base coding to hydrophobic amino acids as they represented in the Hasse diagrams. The Hamming distance between two codons in the Hasse diagram reflects the different hydrophobicities between their respective coded amino acids. Our experiments have demonstrated a small Hamming distance to the wild type HXB2 of almost all the drug-resistant reported mutations in HIV protease gene. The human beta-globin mutant genes have also exhibited similar results. Our research suggests that the Hamming distance between two genes in the molecular evolution process have a minimal value.

[*] Robersy Sánchez: robersy@uclv.edu.cu

Mail address: Apartado postal 697. Santa Clara 1. CP 50100. Villa Clara. Cuba

# 1 INTRODUCTION

So far the rules for encoding base triplets to amino acids in the genetic code order remain an attractive puzzling problem. The genetic code is the biochemical system that establishes the rules to transcribe the nucleotide sequence of a gene into the mRNA codon sequence, later translated into the amino acid sequence of the corresponding protein. The codon set is an extension of the four-letter alphabet found in the DNA molecule. The DNA bases *Adenine*, *Guanine*, *Cytosine*, and *Thymine*, usually denoted *A, G, C, T* (in the RNA, *T* is changed for *U, Uracil*) are paired according to the following rule: *G≡C, A=T* (where each '−' symbolizes a hydrogen bond). It is said that base *G* is the complementary base of *C,* and *A* is that of *T* in the DNA molecule, and vice versa. Moreover, it is well-known that there is an association between the second-position base and hydrophobicity. The amino acids having *U* at the second position of their codons are hydrophobic: {*I, L, M, F, V*} -the amino acids are written using the one-letter symbols and have low polarities- according to the Grantham polarity scale (Grantham, 1974). Those with *A* at the second position, however, are hydrophilic (polar amino acids): {*D, E, H, N, K, Q, Y*} (Crick, 1968). The amino acids with C in the second position of their codons have their polarities in the middle, between the last two groups, while those with *G* in the second position don't follow any regularity in their polarities. The non-random organization of the genetic code has been pointed out (Woese, 1965; Freeland and Hurst,.1998) and various hypothesis have been proposed to explain their enigmatic order whose origin remains unknown. (Woese, 1965; Woese et al. 1966; Crick, 1968; Alf-Steinberger, 1969; Swanson, 1984).

For this purpose, different formal mathematical models of the genetic code have involved the binary representation of the DNA bases. (Jiménez-Montaño et al, 1996; Stambuk, 2000; Karasev and Stefanov 2001; He et al, 2004). Jiménez-Montaño et. al. (1996) suggested a binary interpretation of the genetic code with the following correspondences: *A*= 00, *G* = 01, *U*=10, *C*=11. This binary code of six variables could be represented by the Boolean hypercube. Stambuk (2000), who introduced the universal metric properties of the genetic code, defined it by means of the nucleotide base representation on the square with vertices *U* or *T* = 00, *C* = 01, *G* =10 and *A* = 11. Very differently, Karasev and Stefanov (2001) suggested a model for the topological coding of proteins and arrived at the correspondence

$C$=00, $U$=01, $G$=10, $A$=11. Recently, He et al (2004) used the Gray code representation of the genetic code $C$=00, $U$=10, $G$=11 and $A$=01 to generate a sequence of genetic code-based matrices.

The binary representation of the DNA bases suggests that some partial order should exist both in the base set and in the genetic code. Then, it is normal to think that some partial order in the codon set should reflect the physico-chemical properties of amino acids (Lehmann, 2000; Knight et al (1999).

Our partial order, obtained in the codon set as a consequence of a Boolean lattice and defined in the standard genetic code, reflects the relationship between the codon assignment and the physico-chemical properties of amino acids. Resulting from the partial order of the four DNA bases derived from the Boolean lattices of the four bases, these Boolean lattices are defined by the same physico-chemical properties used by Jiménez-Montaño et al. (1996): hydrogen bondings (to be exact in this paper we use the hydrogen bond number) and the chemical types. Because the number of codons in the genetic code is the number of three-letter variations repeated from the four-letter alphabet, the Boolean lattice on the triplet base set (64-codon set) will be the direct third power of the Boolean lattice of the four DNA bases. Therefore, the aim of this paper is to describe a new Boolean lattice of the standard genetic code and to show its correspondence with the experimental data.

### 1.1 Theoretical Support

This paper's mathematical groundwork is the Boolean lattice. First, the Boolean lattice of the four bases of the DNA is built and from it, the Boolean lattice of the Genetic Code. Actually, the Boolean lattice of the Genetic Code is obtained as the direct third power of the initial lattice. (A definition of a Boolean lattice appears in the Appendix).

Next are the lattice properties used on the set of elements $X$ where the Boolean lattice is denoted $(B(X), \vee, \wedge)$.

- In every Boolean lattice $(B(X), \vee, \wedge)$, for any two $\alpha, \beta \in X$ elements, we have $\alpha \leq \beta$, if and only if $\neg \alpha \vee \beta = 1$. If $\neg \alpha \vee \beta = 1$, it is said that $\beta$ is deduced from $\alpha$. Furthermore, if $\alpha \leq \beta$ or $\alpha \geq \beta$, the elements $\alpha$ and $\beta$ are said to be comparable. Otherwise, they are said not to be comparable and are then denoted $\alpha \parallel \beta$.

- For any Boolean lattice (*B*(*X*), ∨, ∧) there exists the "dual Boolean lattice" (*B'*(*X*), ∧, ∨), where the order relation is reversed, the symbols ∨ and ∧ are interchanged and the maximum and minimum (1 and 0) are inverted. We refer to lattice (*B*(*X*), ∨, ∧) as a primal lattice and to lattice (*B'*(*X*), ∧, ∨) as a dual Boolean lattice.

- Finally, every Boolean lattice has its corresponding directed graph called the Hasse diagram, where two nodes (elements) α and β are connected with a directed edge from α to β (or with a directed edge from β to α) if and only if α≤β (α≥β) and there is no other element between α and β.

- In the Hasse diagram, chains and anti-chains are obtained: A Boolean lattice subset is called a chain, if for any two elements α and β in the subset we have α≤β or α≥β. If any two of its elements are not comparable, the subset is called an anti-chain.

- In the Hasse diagram of the Boolean lattice, the distance function between its nodes is the Hamming distance that indicates the length of the path between two nodes (the number of edges between two nodes.)

## 2   THE BOOLEAN LATTICE MODEL OF THE GENETIC CODE

Following the nature of the partial order of the four DNA bases previously defined by means of the hydrogen bond number and the chemical types of purines {*A, G*} and pyrimidines {*U, C*} bases, the genetic code partial order is built. Actually, codons are base-triplets and the number of codons in the genetic code is the number of three-letter variations with repetition from the four-letter base alphabet.

The Boolean lattice of the four bases is built assuming that the complementary bases in the lattice are the complementary bases in the DNA molecule (*G* ≡ *C* and *A*=*T* or *A*=*U* during the translation of mRNA). That is, the bases with the same number of hydrogen bonds in the DNA molecule and with different chemical types must be complementary elements in the lattice. This lattice, however, needs two non-comparable elements, a maximum element and a minimum element. At this point, the desired Boolean lattice of the genetic code is assumed to be the direct third power of the four-base Boolean lattice. The maximum element in the Boolean lattice of the genetic code has to be the direct third power of the maximum element in the Boolean lattice of the four bases. We then draw the correspondence:

$$U \rightarrow UUU, C \rightarrow CCC, G \rightarrow GGG, A \rightarrow AAA.$$

The Boolean lattice with the desired biological signification has to be selected by taking into account the physico-chemical properties of these codons as well as their respective amino acids. The selection criterion used is supported by the following observations:

1) Both codons *GGG* and *CCC* have the same maximum hydrogen bond numbers and they code for small amino acid side chains with a small polarity difference: *Glycine* and *Proline* respectively (Grantham, 1974). These similarities justify their comparability.

2) Although codons *UUU* and *AAA* have the same minimum hydrogen bond numbers, they code for amino acid side chains with extreme opposite polarities: *Leucine*, a hydrophobic residue and *Lysine* having a strong polar group. This opposite property explains why these elements are not comparable.

These observations allow us to select two dual Boolean lattices, the Primal and the Dual, as they are conventionally known. At first sight, the maximum element in the Primal lattice seemed to be *C* and the minimum *G*, and in the Dual lattice, the maximum element seemed to be *G* and the minimum *C*. The second observation demonstrated that elements *U* and *A* are not comparable and that, therefore, they should not be the maximum or minimum elements in a lattice with biological meaning. So, we have two Boolean lattices $(B(X), \vee, \wedge)$ (the primal lattice) and $(B'(X), \wedge, \vee)$ (the dual lattice), where $X=\{U, C, G, A\}$. The Hasse diagrams of the two Boolean lattices obtained appear in Fig.1. It is obvious that the terms 'primal' and 'dual' in these Boolean lattices are interchangeable but they will not affect the biological meaning. To simplify the notation we will refer simultaneously to both lattices as $B(X)$.

In Fig. 1 the isomorphism of the Boolean lattices $B(X)$ with the Boolean lattices $((Z_2)^2, \vee, \wedge)$ and $((Z_2)^2, \wedge, \vee)$ with $Z_2=\{0,1\}$, is also represented in the model, since all Boolean lattices with the same number of elements are isomorphic. Then, the primal lattice can be represented by means of the correspondence: $G \leftrightarrow 00$; $A \leftrightarrow 01$; $U \leftrightarrow 10$; $C \leftrightarrow 11$. Likewise, for the dual lattice we have: $C \leftrightarrow 00$; $U \leftrightarrow 01$; $A \leftrightarrow 10$; $G \leftrightarrow 11$ (Fig.1).

Two dual genetic code Boolean lattices are obtained from the direct third power of the Boolean lattices $B(X)$ of the four DNA bases: $C(X)=B(X) \times B(X) \times B(X)$.

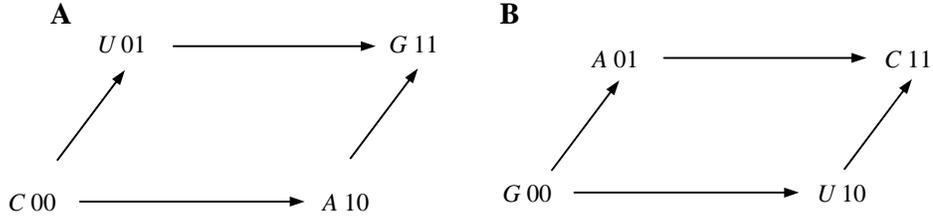

**Figure 1.** The Hasse diagrams of the Boolean lattices. A: The Primal Boolean lattice. B: The Dual Boolean lattice.

Published in Similarly, the *C(X)* Boolean lattices are isomorphic to the dual Boolean lattices $((Z_2)^6, \vee, \wedge)$ and $((Z_2)^6, \wedge, \vee)$. They are induced by the φ: $B(X) \rightarrow (Z_2)^2$ isomorphism, so that, for instance (in the primal lattice):

$GAG \leq AAC \leftrightarrow 000100 \leq 010111$

$ACG \| CGA \leftrightarrow 011100 \| 110001$

$\neg(CAU) = GUA \leftrightarrow \neg(110110) = 001001$

Thus, starting from the genetic code source alphabet consisting of four nucleotides of the DNA and the mRNA, we arrive at the second extension of this alphabet with $2^6=64$ letter-codons of the genetic code. In this Boolean lattice, the distance between two codons using the well-known Hamming distance is figured out. This distance ($d_H$) between two codons, shown as binary sextuplets corresponds to the number of different digits between them. That is,

$d_H(CGU, AUC) = d_H(110010, 011011) = 3$

$d_H(AAG, UGA) = d_H(010100, 100001) = 4$

As mentioned above, this is the distance between the nodes in the Hasse diagram. The Hamming distance between two genes ($D_H$) will be the sum of the Hamming distances between their respective codons. That is, for two genes α and β with *N* codons, we have:

$$D_H(\alpha, \beta) = \sum_{i=1}^{N} d_H(\alpha_i, \beta_i)$$

## 3 RESULTS AND DISCUSSION

The correspondence between the codon order and the physico-chemical properties of amino acids is reflected in the Hasse diagram of the genetic code. This structure, equivalent to a sixth-dimensional Boolean hypercube with vertices representing the codons, is different from those previously reported, (Jiménez-Montaño et al, 1996; Karasev and Stefanov 2001).

Moreover, this ordering of the genetic code shows the natural relevance and strength of the anticodon-codon interaction needed to explain the symmetries in the genetic code table as reflected in our dual Boolean lattices. In both lattices, codons can be read in the 5´→3´ direction and anticodons in the 3´→5´ direction following the standard convention. Consequently, the anticodon of the $^{5'}CUG^{3'}$ codon, represented by 111000 in the primal lattice, is the triplet $^{3'}GAC^{5'}$ similarly represented by 111000 in the dual lattice or represented by 000111 in the primal lattice.

### 3.1 The Hasse Diagram of the Genetic Code

In Fig 2 the Hasse diagram of the primal and the dual lattices simultaneously shows both Boolean lattices reflecting the symmetry-hydrophobicity relationship (the edges of the graph have been left undirected). The symmetric properties of this diagram are determined by the Boolean function *NOT*: $XYZ \rightarrow \neg(XYZ)$, so that if the complementary bases of $X_1$, $X_2$, $X_3$ are the bases $X_1$', $X_2$', $X_3$' ($X_i$, $X'_i \in \{A, C, G, U\}$, $i = 1,2,3$), then the image of codon $^{5'}X_1X_2X_3^{3'}$ is codon $^{5'}X'_1X'_2X'_3^{3'}$. From this last observation, the symmetric image of a codon that codes to hydrophilic amino acids having codons with *A* in the second position is always a codon that codes to hydrophobic amino acids (codons with *U* in the second position). For instance, the symmetric image of the anti-chain {*GUG*, *UGG*, *GGU*, *GGA*, *AGG*, *GAG*} in the Hasse diagram of Fig. 2 is the anti-chain {*CAC*, *ACC*, *CCA*, *CCU*, *UCC*, *CUC*} taking the ordered elements one by one as an image. That is, *GUG* has *CAC* and so on.

All maximal chains have the same length because Boolean lattices are graded. Therefore, each chain from the maximum to the minimum element in both lattices will have the maximum length and vice-versa. In the primal and in the dual lattices all the chains with maximum length will have the same minimum element *GGG* for the primal and *CCC* for the dual and will have the same maximum element *CCC* for the primal and *GGG* for the dual. Moreover, it is evident that two codons will be in the same chain with maximum length if and only if they are comparable, for example, the chain: {*GGG→GAG→AAG→AAA→AAC→CAC→CCC*} (the arrows indicate the deduction direction in the primal lattice) (Fig 2). This diagram has only 720 chains with the same maximum length (each one with six edges).

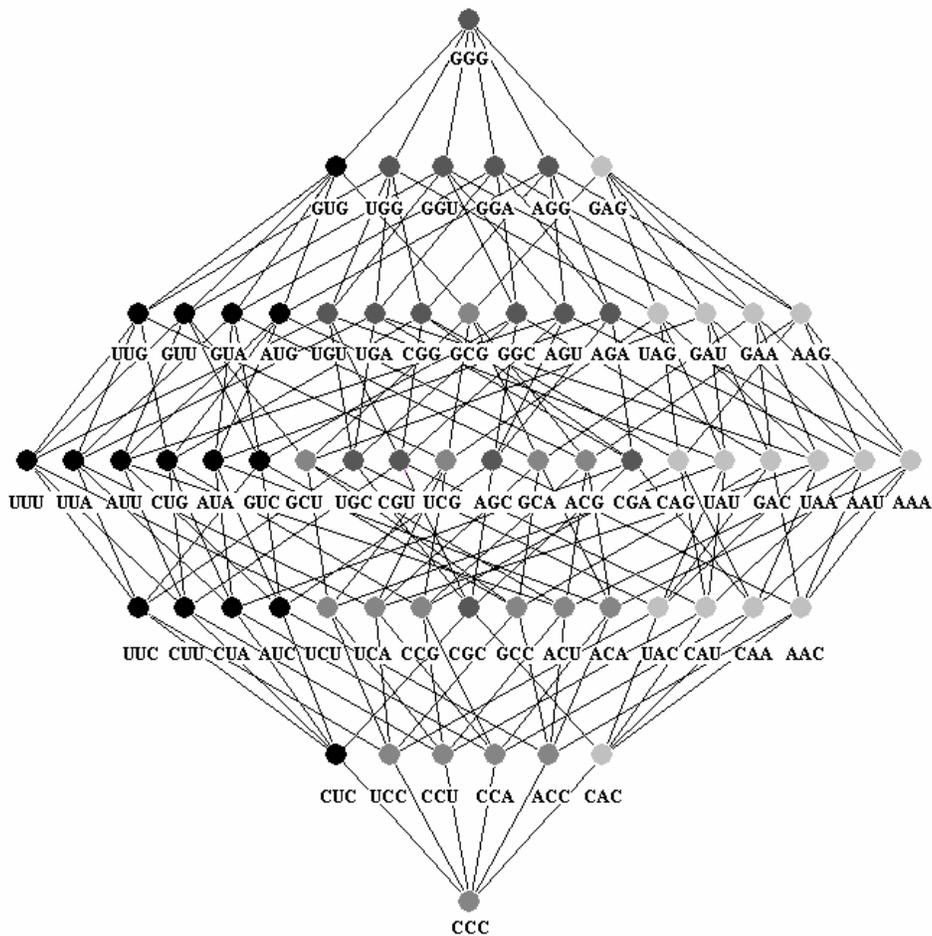

**Figure 2.** The Hasse diagram of the genetic code Boolean lattice. Each grayscale denotes a different group of codons according to the second base. Nodes are black when the second base is *U* coding to hydrophobic amino acids, and dark gray when the second base is *G*. Codons with *C* as a second base are gray, while codons with *A* as a second base coding to hydrophilic amino acids are light gray. Codons *UAA*, *UAG* and *UGA* are terminal codons.

Generally, codons that code to amino acids with extreme hydrophobic differences appear in different chains with maximum length. Particularly, maximal chains with *XUZ* codons (*U* in the second position) lack *X'AZ'* codons. For that reason, it is impossible through deductions to obtain hydrophobic amino acids with codons having *U* in the second position from hydrophilic amino acids with codons having *A* in the second position. These results suggest that the algebraic properties of codons in the Boolean lattices are associated with the hydrophobic properties of amino acids. Kauzmann (1959) expressed that the hydrophobic effects had the main role in the protein process folding (Rose and Wolfenden, 1993) because proteins are surrounded by an aqueous environment and protein-water interactions are considered as the leading power in the folding of the polypeptide chain. Therefore, the algebraic properties of the genetic code Boolean lattices will help us understand the hydrophobic changes in the gene mutation process.

### 3.2 The Hamming Distance and the Physico-Chemical and Biological Properties of Amino Acids

The Hamming distance between two codons in the Hasse diagram reflects the difference between the physico-chemical properties of the corresponding amino acids. In general, between codons *XYZ* and ¬(*XYZ*) (*X, Y, Z* ∈{*A, C, G, U*}) there are larger values of the Hamming distance. Table 1 shows the average of the Hamming distance between the codon sets *XAZ*, *XUZ*, *XCZ* and *XGZ*. The maximum distance corresponds to the transversion in the second base of codons. It is well-known that such transversions are the most dangerous since they frequently alter the hydrophobic properties and the biological functions of proteins.

Specifically, between codons of hydrophilic and hydrophobic amino acids there are great Hamming distance values. Table 2 shows the Hamming distances between amino acid pairs computed as the mean distances between their respective codons.

**Table 1**. Hamming distances between codon subsets *XAZ*, *XUZ*, *XCZ* and *XGZ*. Behind each codon subset, for example, *XUZ* there are 16 realizations. Thus, for every pair of codon subsets there is a symmetric distance matrix with $16^2$ elements. The Hamming distance between two codon subsets is the mean of the 256 distances between their codons. The results are the tabulated integer numbers.

|  | *XGZ* | *XUZ* | *XAZ* | *XCZ* |
|---|---|---|---|---|
| *XGZ* | 2 | 3 | 3 | 4 |
| *XUZ* | 3 | 2 | 4 | 3 |
| *XAZ* | 3 | 4 | 2 | 3 |
| *XCZ* | 4 | 3 | 3 | 2 |

**Table 2**. the Hamming distances between amino acid pairs computed as the mean distances between their respective codons. The Hamming distance values equal or greater than 3.5 appear in bold face.

| * | G | W | C | R | S | V | L | F | M | I | E | D | Y | K | N | Q | H | A | T | P | STOP |
|---|---|---|---|---|---|---|---|---|---|---|---|---|---|---|---|---|---|---|---|---|---|
| G | 1 | 2 | 2 | 2.67 | 3.33 | 2 | **3.67** | 3 | 3 | 3 | 2 | 2 | 3 | 3 | 3 | **4** | **4** | 3 | **4** | **5** | 2.67 |
| W | 2 | 0 | 1.5 | 2.17 | 3.17 | 3 | 2.5 | 2.5 | 3 | **4.33** | 2.5 | **3.5** | 2.5 | **3.5** | **4.5** | 2.5 | **3.5** | **4** | **5** | **4** | 1.33 |
| C | 2 | 1.5 | 0.5 | 2.5 | 2.83 | 3 | 2.83 | 1.5 | **4.5** | **3.83** | **3.5** | 2.5 | 1.5 | **4.5** | **3.5** | **3.5** | 2.5 | **4** | **5** | **4** | 2.17 |
| R | 2.67 | 2.17 | 2.5 | 1.39 | **3.5** | **3.67** | 2.61 | **3.5** | 2.5 | 2.72 | **3.5** | **3.83** | **3.5** | 2.5 | 2.83 | 2.17 | 2.5 | **4.67** | **3.67** | 3.33 | 2.83 |
| S | 3.33 | 3.17 | 2.83 | **3.5** | 2.72 | 3 | 2.94 | 2.5 | **3.5** | 3.28 | 3.17 | 2.83 | 2.5 | **3.5** | 3.17 | 3.17 | 2.83 | 2.67 | 3 | 2.67 | 2.94 |
| V | 2 | 3 | 3 | **3.67** | 3 | 1 | 2.67 | 2 | 2 | 2 | 3 | 3 | **4** | **4** | **4** | **5** | **5** | 2 | 3 | **4** | **3.67** |
| L | **3.67** | 2.5 | 2.83 | 2.61 | 2.94 | 2.67 | 1.39 | 1.83 | 2.17 | 2.39 | **4.5** | **4.83** | **3.83** | **4.17** | **4.5** | 3.17 | **3.5** | **3.67** | 3.33 | 2.33 | 3.17 |
| F | 3 | 2.5 | 1.5 | **3.5** | 2.5 | 2 | 1.83 | 0.5 | **3.5** | 2.83 | **4.5** | **3.5** | 2.5 | **5.5** | **4.5** | **4.5** | **3.5** | 3 | **4** | 3 | 3.17 |
| M | 3 | 3 | **4.5** | 2.5 | **3.5** | 2 | 2.17 | **3.5** | 0 | 1.33 | **3.5** | **4.5** | **5.5** | 2.5 | **3.5** | **3.5** | **4.5** | 3 | 2 | 3 | **4.33** |
| I | 3 | **4.33** | **3.83** | 2.72 | 3.28 | 2 | 2.39 | 2.83 | 1.33 | 0.89 | **4.17** | **3.83** | **4.83** | 3.17 | 2.83 | **4.17** | **3.83** | 3 | 2 | 3 | **4.78** |
| E | 2 | 2.5 | **3.5** | **3.5** | 3.17 | 3 | **4.5** | **4.5** | **3.5** | **4.17** | 0.5 | 1.5 | 2.5 | 1.5 | 2.5 | 2.5 | **3.5** | 2 | 3 | **4** | 1.83 |
| D | 2 | **3.5** | 2.5 | **3.83** | 2.83 | 3 | **4.83** | **3.5** | **4.5** | **3.83** | 1.5 | 0.5 | 1.5 | 2.5 | 1.5 | **3.5** | 2.5 | 2 | 3 | **4** | 2.83 |
| Y | 3 | 2.5 | 1.5 | **3.5** | 2.5 | **4** | **3.83** | 2.5 | **5.5** | **4.83** | 2.5 | 1.5 | 0.5 | **3.5** | 2.5 | 2.5 | 1.5 | 3 | **4** | 3 | 1.83 |
| K | 3 | **3.5** | **4.5** | 2.5 | **3.5** | **4** | **4.17** | **5.5** | 2.5 | 3.17 | 1.5 | 2.5 | **3.5** | 0.5 | 1.5 | 1.5 | 2.5 | 3 | 2 | 3 | 2.83 |
| N | 3 | **4.5** | **3.5** | 2.83 | 3.17 | **4** | **4.5** | **4.5** | **3.5** | 2.83 | 2.5 | 1.5 | 2.5 | 1.5 | 0.5 | 2.5 | 1.5 | 3 | 2 | 3 | **3.83** |
| Q | **4** | 2.5 | **3.5** | 2.17 | 3.17 | **5** | 3.17 | **4.5** | **3.5** | **4.17** | 2.5 | **3.5** | 2.5 | 1.5 | 2.5 | 0.5 | 1.5 | **4** | 3 | 2 | 1.83 |
| H | **4** | **3.5** | 2.5 | 2.5 | 2.83 | **5** | **3.5** | **3.5** | **4.5** | **3.83** | **3.5** | 2.5 | 1.5 | 2.5 | 1.5 | 1.5 | 0.5 | **4** | 3 | 2 | 2.83 |
| A | 3 | **4** | **4** | **4.67** | 2.67 | 2 | **3.67** | 3 | 3 | 3 | 2 | 2 | 3 | 3 | 3 | **4** | **4** | 1 | 2 | 3 | 3.33 |
| T | **4** | **5** | **5** | **3.67** | 3 | 3 | 3.33 | **4** | 2 | 2 | 3 | 3 | **4** | 2 | 2 | 3 | 3 | 2 | 1 | 2 | **4.33** |
| P | **5** | **4** | **4** | 3.33 | 2.67 | **4** | 2.33 | 3 | 3 | 3 | **4** | **4** | 3 | 3 | 3 | 2 | 2 | 3 | 2 | 1 | 3.33 |
| STOP | 2.67 | 1.33 | 2.17 | 2.83 | 2.94 | **3.67** | 3.17 | 3.17 | **4.33** | **4.78** | 1.83 | 2.83 | 1.83 | 2.83 | **3.83** | 1.83 | 2.83 | 3.33 | **4.33** | 3.33 | 0.89 |

*Amino acids are represented in one-letter symbols.

Generally, we can notice that amino acids with large differences in their physico-chemical properties have high Hamming distance values. For instance, the mean distances between the Tryptophan codon and the codons of Isoleucine, Alanine and Threonine are greater than 4.

The tendency of the Hamming distance between amino acids to increase along with their physico-chemical differences is shown by taking the amino acid Euclidean distances represented as three variable vectors: the polarizability parameter (Charton and Charton, 1982), the polarity (Grantham, 1974) and the Normalized van der Waals volume (Fauchere et al., 1988). Next, the 20x20 Euclidean distance matrix between amino acids is calculated. For each codon subset (*XUZ, XCZ, XGZ, XAZ*) two $n$x20 sub-matrices are taken (*n* is the number of amino acids coded in the subset): one sub-matrix from the Euclidean distance matrix and the other from the Hamming distance matrix (presented in Table 2). For instance, for the codon subset *XUZ* $n$=5 (*F, I, M, L, V*) the correlation coefficient between the two 5x20 sub-matrices (Euclidean versus Hamming distances) is equal to 0.59. The correlation coefficient between the corresponding 7x20 sub-matrices of the codon subset *XAZ* (*D, E, H, K, N, Q, Y*) is equal to 0.56 and the correlation coefficient between the corresponding 4x20 sub-matrices of the codon subset *XCZ* (*A, P, S, T*) is equal to 0.44. All these correlations are significant at the 0.001 level (2-tailed test). In the 5x20 sub-matrices corresponding to the codon subset *XGZ* (*C, G, R, S, W*) the correlation is very small and not significant. Fig 3 shows the graphics of the Hamming distance vs. the Euclidean distance from the 7x20 and 5x20 sub-matrices resulting from the subsets *XAZ* and *XUZ*. Even though these correlation coefficients are small, the tendency is clear and highly significant.

Hence, we expect to find the most frequent mutations in genes having small Hamming distances. This idea has been verified experimentally in mutant genes. Table 3 shows the Hamming distance between single point drug resistance mutations in the HIV-1 protease gene and the respective gene of the HXB2 strain. It can be observed that most of the single point mutations have a Hamming distance equal to or lower than 2. A similar situation is found in the reported mutant sequences of the human beta-globin gene (Table 4). It can be seen that small changes in the physico-chemical properties of the amino acids are sufficient to alter the biological function of hemoglobin.

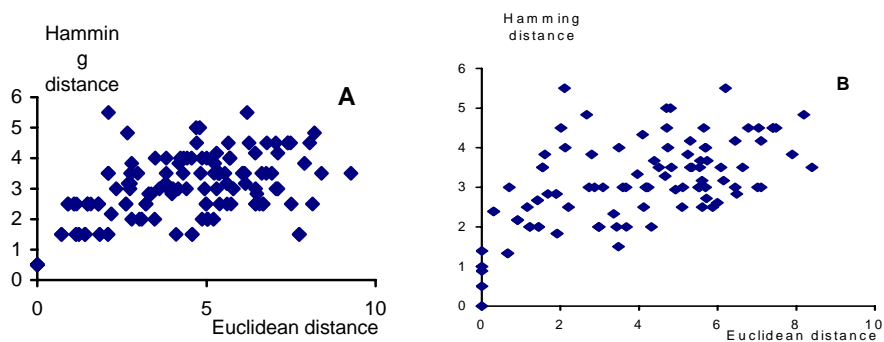

**Figure 3**. The graphics of the Euclidean distance vs. the Hamming distance for amino acids. We have computed the mean of the Hamming distance between the respective codons (see Table 1) for each amino acid pair. The Euclidean distance between each amino acid pair was calculated from its representation as a vector of three variables: polarizability parameter (Charton and Charton, 1982), polarity (Grantham, 1974) and normalized van der Waals volume (Fauchere et al., 1988). Graphic *A* was obtained from 7x20 distance sub-matrices corresponding to hydrophilic amino acids with codons *XAZ*: *D, E, H, K, N, Q, Y*. Graphic *B* was obtained from 5x20 distance sub-matrices corresponding to hydrophobic amino acids with codons *XUZ*: *F, I, M, L, V*. The correlation coefficients are 0.56 and 0.59 respectively, and both are significant at level 0.001 (2-tailed test).

**Table 3.** The Hamming distance of the mutations found in the HIV protease gene that confers drug resistance with regard to the wild type of HXB2. Almost all the reported mutations in the HIV protease gene have a Hamming distance equal to or less than 2. Mutations with the Hamming distances greater than 2 are presented in bold face.

| *Amino acid Mutations | Codon mutation | $d_H$ | Antiviral drug | Amino acid Mutations | Codon mutation | $d_H$ | Antiviral drug |
|---|---|---|---|---|---|---|---|
| A71I | GCU-->AUU | 2 | ABT-378 | **L10Y** | **CUC-->UAC** | **3** | BMS 232632 |
| **A71L** | **GCU-->CUC** | **4** | ABT-378 | L23I | CUA-->AUA | 1 | BILA 2185 BS |
| A71T | GCU-->ACU | 1 | Indinavir, Crixivan | L24I | UUA-->AUA | 2 | Indinavir, Crixivan |
| A71V | GCU -->GUU | 1 | Nelfinavir, Viracept | L24V | UUA-->GUA | 1 | Telinavir |
| D30N | GAU-->AAU | 1 | Nelfinavir, Viracept | L33F | UUA-->UUC | 1 | ABT-538 |
| D60E | GAU-->GAA | 2 | DMP 450 | L63P | CUC-->CCC | 1 | ABT-378, AG1343 |
| G16E | GGG-->GAG | 1 | ABT-378 | L90M | UUG -->AUG | 2 | Nelfinavir, Viracept |
| G48V | GGG -->GUG | 1 | Telinavir, MK-639 | L97V | UUA-->GUA | 1 | DMP-323 |
| G52S | GGU-->AGU | 1 | AG1343 | M36I | AUG-->AUA | 1 | Nelfinavir, Viracept |
| G73S | GGU-->AGU | 1 | AG1343 MK-639 | **M46F** | **AUG-->UUC** | **4** | A-77009 |
| H69Y | CAU-->UAU | 1 | Aluviran, Lopinavir | M46I | AUG-->AUA | 1 | Indinavir, Crixivan |
| I47V | AUA-->GUA | 1 | ABT-378 | **M46L** | **AUG-->UUC** | **4** | Indinavir, Crixivan |
| I50L | AUU-->CUU | 1 | BMS 232632 | M46V | AUG-->GUG | 1 | A-77006 |
| I54L | AUC-->CUC | 1 | ABT-378 | N88D | AAU-->GAU | 1 | Nelfinavir, Viracept |
| I54M | AUU-->AUG | 1 | BILA 2185 BS | N88S | AAU-->AGU | 1 | BMS 232632 |
| I54T | AUC-->ACC | 1 | ABT-378 | P81T | CCU-->ACU | 1 | Telinavir |
| I54V | AUC-->GUC | 1 | ABT-378, MK-639 | R8K | CGA-->AAA | 2 | A-77003 |
| I82T | AUC-->ACC | 1 | A-77003 | R8Q | CGA-->CAA | 1 | A-77004 |
| I84A | AUA-->GCA | 2 | BILA 1906 BS | R57K | AGA-->AAA | 1 | AG1343 |
| I84V | AUA-->GUA | 1 | Nelfinavir, Viracept | T91S | ACU-->UCU | 2 | ABT-378 |
| K20M | AAG -->AUG | 2 | Indinavir, Crixivan | V32I | GUA-->AUA | 1 | A-77005,Telinavir |
| K20R | AAG-->AGG | 1 | Indinavir, Crixivan | V75I | GUA-->AUA | 1 | Telinavir |
| K45I | AAA-->AUA | 2 | DMP-323 | V77I | GUA-->AUA | 1 | AG1343 |
| K55R | AAA-->AGA | 1 | AG1343 | V82A | GUC-->GCC | 1 | Ritonovir, Norvir |
| L10I | CUC-->AUC | 1 | Indinavir, Crixivan | V82F | GUC-->UUC | 1 | Ritonovir, Norvir |
| L10R | CUC-->CGC | 1 | Indinavir, Crixivan | V82I | GUC-->AUC | 1 | A-77011 |
| L10V | CUC-->GUC | 2 | Indinavir, Crixivan | V82S | GUC-->UCC | 2 | Ritonovir, Norvir |
| L10F | CUC-->UUC | 1 | Lopinavir, | V82T | GUC-->ACC | 2 | Ritonovir, Norvir |

* All mutation information in this table was taken from Los Alamos web site: http://resdb.lanl.gov/Resist_DB

For instance, mutations *V*20*M*, *D*21*N*, *V*121*L* and *D*99*E* preserve the hydrophobic character but alter the oxygen affinity to hemoglobin. Such experimental results suggest that the mutational pathways followed by the genes in the molecular evolution process tend to have the smallest Hamming distance in each step. In fact, this hypothesis might be called 'continuity hypothesis' and this suggest that small differences in the biological activity between the wild type and the mutant gene would mean a small Hamming distance.

**Table 4**. The Hamming distance of the mutations found in the human beta-globin gene.

| *Amino acid Mutations | Codon mutation | $d_H$ | Biological effect | Reference [PMID**] |
|---|---|---|---|---|
| P36H | CCT-->CAT | 1 | High oxygen affinity | [11939509] Hemoglobin. 2002, 26,:21-31 |
| T123I | ACC-->ATC | 1 | Asymptomatic | [11300351] Hemoglobin. 2001, 25, 67-78. |
| V20E | GTG-->GAG | 2 | High oxygen affinity | [7914875] Eur J Haematol. 1994, 53, 21-5 |
| V20M | GTG-->ATG | 1 | High oxygen affinity | [7914875] Eur J Haematol. 1994, 53, 21-5 |
| V126L | GTG-->CTG | 2 | Neutral | [11939515] Hemoglobin. 2002, 26, 7-12 |
| V111F | GTC-->TTC | 1 | Low oxygen affinity | [10975442] Hemoglobin. 2000, 24, 227-37 |
| H97Q | CAC-->CAA | 1 | High oxygen affinity | [8571935] Am J Hematol. 1996, 51, 32-6 |
| V34F | GTC-->TTC | 1 | High oxygen affinity | [10846826] Int J Hematol. 2000,71, 221-6 |
| E121Q | GAA-->CAA | 2 |  | [8095930] Hemoglobin. 1993, 17, 9-17 |
| L114P | CTG-->CCG | 1 | Non-functional | [11300352] Hemoglobin. 2001, 25, 79-89 |
| A128V | GCT-->GTT | 1 | Mild instability | [11300349] Hemoglobin. 2001, 25, 45-56 |
| H97Q | CAC-->CAG | 2 | High oxygen affinity | [8890707] Ann Hematol.1996, 73,183-8 |
| D99E | GAT-->GAA | 2 | High oxygen affinity | [1814856] Hemoglobin. 1991, 15, 487-96 |
| D21N | GAT-->AAT | 1 |  | [8507722] Ann Hematol. 1993, 66, 269-72 |
| N139Y | AAT-->TAT | 2 | High oxygen affinity | [8718692] Hemoglobin. 1995, 19, 335-41 |
| V34D | GTC-->GAC | 2 | Unstable | [1260309] Hemoglobin. 2003, 27, 31-5 |
| E121K | GAA-->AAA | 1 |  | [7908281] Hemoglobin. 1993, 17, 523-35 |
| A140V | GCC-->GTC | 1 | Mild polycythemia | [9028820] Hemoglobin. 1997, 21, 17-26 |
| K82E | AAG-->GAG | 1 | Altered oxygen affinity | [9255613] Hemoglobin. 1997; 21, 345-61 |
| G83D | GGC-->GAC | 1 | Hb Pyrgos (Normal) | [11843288] Int J Hematol. 2002, 75, 35-9 |
| D99N | GAT-->AAT | 1 | High oxygen affinity | [1427427] Haematologica. 1992, 77,:215-20 |
| G15R | GGT-->CGT | 2 | Neutral | [11939517] Hemoglobin. 2002, 26, 77-81 |
| V111L | GTC-->CTC | 2 | Fannin-Lubbock variant | [7852084] Hemoglobin. 1994, 18, 297-306 |
| G119D | GGC-->GAC | 1 | Fannin-Lubbock variant | [7852084] Hemoglobin. 1994, 18, 297-306 |
| E26K | GAG-->AAG | 1 |  | [9140717] Hemoglobin. 1997, 21, 205-18 |
| N108I | AAC-->ATC | 2 | Low affinity | [12010673] Haematologica. 2002, 87, 553-4 |
| H146P | CAC-->CCC | 1 | High oxygen affinity | [11475152] Ann Hematol. 2001, 80, 365-7 |
| H92Y | CAC-->TAC | 1 | Cyanosis | [9494043] Hemoglobin. 1998, 22, 1-10 |
| C112W | TGT-->TGG | 1 | Silent and unstable | [8936462] Hemoglobin. 1996, 20, 361-9 |
| A111V | GCC->GTC | 1 | Silent | [7615398] Hemoglobin. 1995,19, 1-6 |
| A123S | GCC-->TCC | 1 | Silent | [7615398] Hemoglobin. 1995,19, 1-6 |
| D52G | GAT-->GGT | 1 | Silent | [9730366] Hemoglobin. 1998, 22, 355-71 |
| V126G | GTG-->GGG | 1 | Mild beta-thalassaemia | [1954392] Blood. 1991,78, 3070-5 |
| W15Stop | TGG-->TAG | 1 | Beta-thalassaemia | [10722110] Hemoglobin. 2000 Feb;24(1):1-13 |
| F42L | TTT-->TTG | 1 | Hemolytic anemia | [11920235] Hematol J. 2001;2(1):61-6 |

* Amino acids are represented by one-letter symbols. **Numbers are the PubMed ID of the articles indexed by MEDLINE where the mutations were reported.

The 'continuity hypothesis' has been verified by reported mutation experiments with the HIV-1 protease and the reverse-transcriptase (Rose R. E. et al., 1996, Kim et al., 1996). Fig 4 *A* shows the fold increases compared to the wild type of resistant mutations on HIV-1 protease against the protease inhibitor MK-639.

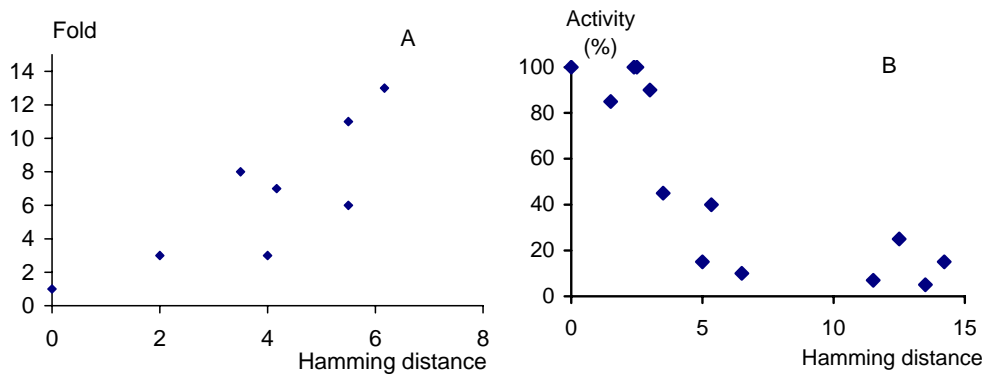

**Figure 4**. Biochemical activity vs. the Hamming distance. *A*: The fold increases compared to the wild type of the resistant mutations on HIV-1 protease against the protease inhibitor MK-639 (Rose et al., 1996) and *B*: The enzymatic activity changes in the HIV-1 reverse-transcriptase mutants normalized with regard to the wild type enzyme (Kim et al., 1996).

Fig 4 *B* shows the graphs of the reverse-transcriptase activity, normalized with regard to the wild type, versus the Hamming distance. Generally, it can be observed that a small Hamming distance between the wild type and the mutant means a small difference in their biochemical activities.

Such results support the assumption that the genetic code reduces the effects of point mutations and minimizes the subsequent transcription and translation errors to make the reproduction of genetic information possible (Friedman and Weinstein, 1964; Epstein, 1966; Crick, 1968; Alf-Steinberger, 1969; Parker, 1989; Gillis et al. 2001). These are the results of the genetic code order. The arrangement of codons in the genetic code is such that the Hamming distance, which is determined by the physico-chemical properties of the four DNA bases, is connected with the physico-chemical properties of amino acids.

## 4 CONCLUSIONS

The Boolean lattices obtained reflect a strong connection between the genetic code order and the physico-chemical properties of amino acids. Such connections are evident in the Hasse diagrams of the Boolean lattices. The symmetric image of a codon with *U* as a second base codifying to hydrophobic amino acids is always a codon with *A* as a second base codifying to hydrophilic amino acids. Moreover, the transversions at the second codon position have the largest Hamming distance average. Specifically, the average of the Hamming distance between the codon sets *XUZ* and *X'AZ'* is the maximum.

The results of these assumptions have been verified in the experimental data. Almost all the drug resistance reported mutations in HIV protease gene have a small Hamming distance with regard to the wild type, determined in the Boolean lattices. Likewise, the single point mutations in the mutants of the human beta-globin genes have a small Hamming distance regarding their wild types. Furthermore, we found that the small difference between the enzyme activities of the wild type and the mutant means a small Hamming distance between them. The experimental confrontation suggests that in the molecular evolution process, the mutation pathway tends to have the minimal Hamming distance between the wild type and the mutant genes (proteins) in each mutation step. These results advance the idea that the Boolean lattice could allow us to model the gene mutation process.

APPENDIX

## The Boolean Lattice Definition

We would like to go over some definitions of the Boolean lattice for the advantage of the reader. The lattice concept is connected with the concept of a partially ordered set.

**Definition:** A partially ordered set $X$ is a set of elements with a binary relation, denoted by "$\leq$", which is:

   i. Reflexive: $\alpha \leq \alpha$ for all $\alpha \in X$

   ii. Transitive: $\alpha \leq \beta$ and $\beta \leq \delta$ imply $\alpha \leq \delta$

   iii. Anti-symmetric: $\alpha \leq \beta$ and $\beta \leq \alpha$ imply $\alpha = \beta$

If for $x \in X$ the element $\alpha$ satisfies the inequality $\alpha \leq x$ then, the element $\alpha$ is called a "lower bound" to the set $X$. The lower bound $\beta$ is called the greatest lower bound (g.l.b) if $\beta \geq \alpha$ for any other lower bound $\alpha$. Dually, we can defined "upper bound" and "least upper bound" (l.u.b.)(Birkhoff and MacLane, 1941). The g.l.b. of $\alpha$ and $\beta$ is denoted by $\alpha \vee \beta$ and the l.u.b. by $\alpha \wedge \beta$.

**Definition:** A Boolean lattice $B(X)$ is a partially ordered set of elements with the following properties:

1. $B(X)$ contains two elements called the minimum and maximum elements, denoted by 0 and 1 respectively, which are universal bounds, that is: $0 \leq \alpha \leq 1$ for all $\alpha \in X$ and satisfies the special properties:

   i. Intersection: $0 \wedge \alpha = 0$ and $1 \wedge \alpha = \alpha$

   ii. Union: $0 \vee \alpha = \alpha$ and $1 \vee \alpha = 1$

2. For all elements $\alpha \in X$ there is the element $\neg \alpha \in X$, called complement of the element $\alpha$, such that:

$$\alpha \wedge \neg \alpha = 0 \text{ and } \alpha \vee \neg \alpha = 1$$

3. In $B(X)$ the operations $\wedge$ and $\vee$ satisfy the distributive law, that is:

$$(\alpha \wedge \beta) \vee (\alpha \wedge \delta) = \alpha \wedge (\beta \vee \delta) \text{ and}$$

$$(\alpha \vee \beta) \wedge (\alpha \vee \delta) = \alpha \vee (\beta \wedge \delta)$$